\documentclass[12pt,prd,tightenlines,nofootinbib,showpacs,showkeys]{revtex4}
\newcommand{\be}{\begin{equation}}
\newcommand{\ee}{\end{equation}}
\newcommand{\bdis}{\begin{displaymath}}
\newcommand{\edis}{\end{displaymath}}
\newcommand{\bga}{\begin{equation}\begin{gathered}}
\newcommand{\ega}{\end{gathered}\end{equation}}
\usepackage{bm}
\usepackage{graphics}
\usepackage{rotating}
\usepackage{epsfig}
\usepackage{amssymb}
\usepackage{amsmath}
\usepackage{amsthm}
\usepackage{amsfonts}
\usepackage{amssymb}
\usepackage{braket}
\usepackage{graphics}
\usepackage{rotating}
\usepackage{epsfig}
\usepackage{indentfirst}
\usepackage{epstopdf}

\begin{document}
\title{\begin{flushright}{\rm\normalsize SSU-HEP-15/08\\[5mm]}\end{flushright}
Hyperfine structure of P-states in muonic deuterium}
\author{\firstname{R.~N.} \surname{Faustov}}
\affiliation{Dorodnicyn Computing Centre, Russian Academy of Science, Vavilov Str. 40, 119991, Moscow, Russia}
\author{\firstname{A.~P.} \surname{Martynenko}}
\affiliation{Samara State University, Pavlov Str. 1, 443011, Samara, Russia}
\affiliation{Samara State Aerospace University named after S.P. Korolyov, Moskovskoye Shosse 34, 443086,
Samara, Russia}
\author{\firstname{G.~A.} \surname{Martynenko}}
\author{\firstname{V.~V.} \surname{Sorokin}}
\affiliation{Samara State University, Pavlov Str. 1, 443011, Samara, Russia }

\begin{abstract}
On the basis of quasipotential approach to the bound state problem in quantum electrodynamics
we calculate hyperfine structure intervals $\Delta E^{hfs}(2P_{1/2})$ and $\Delta E^{hfs}(2P_{3/2})$
for P-states in muonic deuterium. The tensor method of projection operators for the calculation
of the hyperfine structure of P-states with definite quantum numbers of total atomic momentum $F$
and total muon momentum $j$ in muonic deuterium is formulated.
We take into account vacuum polarization, relativistic,
quadrupole and structure corrections of orders $\alpha^4$, $\alpha^5$ and $\alpha^6$.
The obtained numerical values of hyperfine splittings are useful for the analysis of new experimental data
of the CREMA collaboration regarding to muonic deuterium.
\end{abstract}



\maketitle

\section{Introduction}

The investigation of energy spectrum of light muonic atoms (muonic hydrogen, muonic deuterium,
ions of muonic helium) reached a new level at present. This is due to new experimental results obtained
by the CREMA collaboration in \cite{CREMA1,CREMA2015,CREMA2}. On the one side these results open a possibility to
obtain new values of a number of fundamental physical constants such as nuclear charge radii.
While experimental data on atomic transitions have become very precise, our knowledge of the charge radii,
which are part of theoretical predictions, is not as accurate as we would like.
On the other hand, they call to look again at the formulation of the theory of bound states in quantum
electrodynamics and possibly revise some of its previous aspects. The second position was proved important after
a series of experiments in \cite{CREMA1,CREMA2} which revealed essential disagreement between two values of the
proton charge radius obtained in experiments with electronic and muonic atoms \cite{CREMA1,CREMA2,egs}.
An analysis of the situation and determining the causes of discrepancies are investigated in several directions,
which are widely discussed in \cite{CREMA2015,str,str1,str2,barger,str3,pi,sgk,str4,str5}. It is possible that the publication of
new experimental data on the structure of the energy levels of muonic deuterium which is planned in near future,
will help clarify the problem. A comparison of the theory and experiment for the transition
frequencies $\nu (2^{2F+1}P_j\div 2^{2F'+1}S_{j'})$ in muonic deuterium demands careful consideration of different contributions
to the energy P-levels. The calculations of fine and hyperfine structure of the energy spectrum of light muonic
atoms were made in a series of papers \cite{borie1,borie2,borie3}. The results of these studies are a reliable
benchmark for a comparison with experimental data and provide a starting point for further research.
Whereas the calculation of separate contributions to the hyperfine structure of S-states of the muonic deuterium,
even with a very specific kind, was the subject of intense study, the hyperfine structure of P-states much less investigated.
Therefore, in this study we aim to partly fill this gap.
In this work we make new analysis of different corrections to hyperfine splittings
of P-states which allow to obtain more accurate results important for a comparison with experimental data.
Another aim of our study is to develop a method of projection operators in the investigation of the energy structure
of P-states. The method of projection operators on the bound states with definite spins was used previously in
\cite{fmms,fmms1} for the construction of particle interaction operator for hyperfine structure of S-states.

\section{General formalism}

Let us begin our consideration with basic contributions to hyperfine structure of P-states of order $\alpha^4$.
Our approach to the calculation of hyperfine splittings is based on quasipotential method in quantum
electrodynamics in which the two-particle bound state is described by the Schr\"odinger equation \cite{fm2004,apm2005,apm2008}.
In this work we develop another approach to the calculation of hyperfine structure of muonic deuterium based on tensor
representation of P-wave projection operators describing muonic deuterium states. First we show on an example of calculating
the leading order contributions how a tensor formalism helps investigate the hyperfine structure of the spectrum.
It is useful to work in momentum representation where we can write
the wave function of muonic deuterium 2P-state in the tensor form:
\begin{equation}
\label{eq:q1}
\psi_{2P}({\bf p})=\left(\varepsilon\cdot n_p\right)R_{21}(p),
\end{equation}
where $\varepsilon_\delta$ is the polarization vector of orbital motion, $n_p=(0,{\bf p}/p)$, $R_{21}(p)$ is
the radial wave function in momentum space. Then the energy shifts are presented in integral form:
\begin{equation}
\label{eq:q2}
\Delta E^{hfs}=\int\left(\varepsilon^\ast\cdot n_q\right)R_{21}(q)\frac{d{\bf q}}{(2\pi)^{3/2}}
\int\left(\varepsilon\cdot n_p\right)R_{21}(p)\frac{d{\bf p}}{(2\pi)^{3/2}} \Delta V^{hfs}({\bf p},{\bf q}).
\end{equation}
In the leading order the hyperfine potential $\Delta V^{hfs}$ is constructed by means of one-photon interaction amplitude $T_{1\gamma}$ .
Writing  the amplitude $T_{1\gamma}$ we refer to it a part of the bound state wave function related to orbital motion:
\begin{equation}
\label{eq:q3}
T_{1\gamma}({\bf p},{\bf q})=4\pi Z\alpha \left(\varepsilon^\ast\cdot n_q\right)\left[\bar u(q_1)
\left(\frac{p_{1,\mu}+q_{1,\mu}}{2m_1}+(1+a_\mu)\sigma_{\mu\epsilon}\frac{k_\epsilon}{2m_1}\right)u(p_1)\right]
\left(\varepsilon\cdot n_p\right)D_{\mu\nu}(k)\times
\end{equation}
\begin{displaymath}
\varepsilon^\ast_{d,\rho}(q_2)\Bigl\{g_{\rho\sigma}\frac{(p_2+q_2)_\nu}{2m_2}F_1(k^2)-\frac{(p_2+q_2)_\nu}
{2m_2}\frac{k_\rho k_\sigma}{2m_2^2}F_2(k^2)+
(g_{\rho\lambda}g_{\sigma\mu}-g_{\rho\mu}g_{\sigma\lambda})\frac{k_\lambda}{2m_2}F_3(k^2)\Bigr\}\varepsilon_{d,\sigma}(p_2),
\end{displaymath}
where $p_{1,2}=\frac{m_{1,2}}{(m_1+m_2)}P\pm p$ are four-momenta of initial muon and deuteron, $q_{1,2}=
\frac{m_{1,2}}{(m_1+m_2)}Q\pm q$
are four-momenta of final muon and deuteron. They are expressed in terms of total two-particle momenta $P,Q$
and relative momenta $p,q$.
$D_{\mu\nu}(k)$ is the photon propagator which is taken to be in the Coulomb gauge.
Explicit expression of the deuteron wave function $\varepsilon_{d,\sigma}(p)$
has the form:
\begin{equation}
\label{eq:p1}
\varepsilon_{d,\sigma}(p_2)=\varepsilon_{d,\sigma}(0)-\frac{p_{2,\sigma}+g_{0\sigma}m_2}{\epsilon_2(p)+m_2}\frac{(\varepsilon_{d,\sigma}(0)\cdot p_2)}{m_2}.
\end{equation}
It should be noted that the amplitude \eqref{eq:q3} has been studied in detail
in \cite{brodsky} excepting quadrupole correction. In the center-of-mass rest frame $P=Q=Mv$, $v=(1,0)$.
The form factors $F_{1,2,3}(k^2)$ are related to the
charge, magnetic and quadrupole deuteron form factors as ($\eta =k^2/4m_2^2$) \cite{abbott,apm2003}:
\begin{equation}
\label{eq:fcfmfq}
F_C=F_1+\frac{2}{3}\eta\left[F_1+(1+\eta)F_2-F_3\right],~~~F_M=F_3,~~~F_Q=F_1+(1+\eta)F_2-F_3.
\end{equation}
We consider \eqref{eq:q3} as a starting point for a composition
of orbital ${\bf L}$ momentum, the deuteron spin ${\bf s}_2$ (note that the spin of the nucleus is usually denoted by {\bf I})
and muon spin ${\bf s}_1$. In the first scheme of momentum composition
we add firstly momenta ${\bf L}$ and ${\bf s}_1$ obtaining two muon states with angular momenta $j=1/2$ and $j=3/2$. In the Rarita-Schwinger
formalism the wave function of the state with half-integer spin $3/2$ is described by
\begin{equation}
\label{eq:q4}
\psi_{\mu,\alpha}({\bf p},\sigma)=\sum_{\lambda,\omega}\Braket{\frac{1}{2}\omega;1\lambda|\frac{3}{2}\sigma}\varepsilon_\mu({\bf p},\lambda)
u_\alpha({\bf p},\omega),
\end{equation}
where $\Braket{\frac{1}{2}\omega;1\lambda|\frac{3}{2}\sigma}$ are the Clebsch-Gordon coefficients.
Another sequence of angular momentum addition is that in the beginning we add the orbital and intrinsic angular momentum of the deuteron
and then the muon spin.
When we combine the $L=1$ and $s_2=1$ we get three states with the deuteron momenta $2,1,0$. The deuteron wave function has in
this case the form:
\begin{equation}
\label{eq:q5}
\phi_{\mu\nu}({\bf p},\gamma)=\sum_{\lambda_1,\lambda_2}\Braket{1\lambda_1;1\lambda_2|2\gamma}\varepsilon_\mu({\bf p},\lambda_1)
\varepsilon_\nu({\bf p},\lambda_1).
\end{equation}
After combining $\phi_{\mu\nu}$ with the muon spin on the second stage there arise three states with $F=5/2, 3/2, 1/2$
which are described by the tensor-spinor field $\Psi_{\mu\nu}$ satisfying to the Dirac equation
\begin{equation}
\label{eq:q6}
(\hat v-1)\Psi_{\mu\nu}=0,~~~v^\mu\psi_{\mu\nu}=0.
\end{equation}
The field $\Psi_{\mu\nu}$ can be easily decomposed into different parts with definite atomic angular momentum $F$:
\begin{equation}
\label{eq:q7}
F^P=\frac{5}{2}^-:~~~\Psi_{\mu\nu};
\end{equation}
\begin{equation}
\label{eq:q8}
F^P=\frac{3}{2}^-:~~~\Psi^S_{\mu\nu}=\frac{1}{\sqrt{10}}\left(\gamma_{\perp\mu}\gamma_5\psi_\nu+\gamma_{\perp\nu}\gamma_5\psi_\mu\right),
\end{equation}
\begin{equation}
\label{eq:q9}
F^P=\frac{3}{2}^-:~~~\Psi^A_{\mu\nu}=\frac{1}{\sqrt{2}}\left(\gamma_{\perp\mu}\gamma_5\psi_\nu-\gamma_{\perp\nu}\gamma_5\psi_\mu\right),
\end{equation}
\begin{equation}
\label{eq:q10}
F^P=\frac{1}{2}^-:~~~\Psi^A_{\mu\nu}=\frac{1}{2\sqrt{6}}\left[\gamma_{\perp\mu},\gamma_{\perp\nu}\right],
\end{equation}
\begin{equation}
\label{eq:q11}
F^P=\frac{1}{2}^-:~~~\Psi^S_{\mu\nu}=\frac{1}{\sqrt{3}}\left(g_{\mu\nu}-v_\mu v_\nu\right),
\end{equation}
where $\Psi_{\mu\nu}$ is the usual $5/2$ generalized, symmetric Rarita-Schwinger tensor-spinor \cite{berends,mosel}.
The negative parity is obvious from physical reasons. Different states with
total momentum $F=1/2$ and $F=3/2$ are decomposed into symmetric and antisymmetric parts satisfying to \eqref{eq:q6}.
The tensor-spinor wave functions were used previously in \cite{koerner} for the bound states of quarks.
For further calculations, we note that each field $\Psi_{\mu\nu}^{S,A}$ with $F=3/2,1/2$ is a superposition of states with
muon angular momentum $j=1/2$ and $j=3/2$. Introduced in \eqref{eq:q7}-\eqref{eq:q11} tensor-spinor fields can be considered
as both a projection operators on the states with a definite value of the total angular momentum.
These projectors are very convenient for the calculation of the matrix elements of the interaction potential corresponding
to certain quantum numbers. They allow us to avoid direct cumbersome multiplication of different factors
in the amplitudes of the interaction of particles and use the computer methods for calculating amplitudes and the energy shifts
\cite{form}.

To demonstrate this property of $\Psi_{\mu\nu}$ we continue our calculations of the amplitude \eqref{eq:q3} corresponding
to transitions between states with definite values of $F$. Introducing projectors $\Psi_{\mu\nu}$ in \eqref{eq:q3} and averaging
the amplitude over the projection of the total angular momentum ${\cal M}$ we obtain the following basic relation:
\begin{equation}
\label{eq:q12}
\overline{T_{1\gamma}({\bf p},{\bf q})}=\frac{4\pi Z\alpha}{2F+1} n_q^\delta n_p^\omega
Tr\Biggl\{\left[\sum_{{\cal M}=-F}^{F} \Psi^{\cal M}_{\omega\sigma_1}\bar\Psi^{\cal M}_{\delta\rho_1}\right]
\frac{[(m_1(\hat v+1)-{\boldsymbol\gamma}{\bf q}]}{2m_1}\Gamma_\mu\frac{[m_1(\hat v+1)-{\boldsymbol\gamma}{\bf p}]}{2m_1}\Biggr\}\times
\end{equation}
\begin{equation*}
\Bigl\{g_{\rho\sigma}\frac{(p_2+q_2)_\nu}{2m_2}F_1(k^2)-\frac{(p_2+q_2)_\nu}{2m_2}\frac{k_\rho k_\sigma}{2m_2^2}F_2(k^2)+(g_{\rho\lambda}g_{\sigma\mu}-g_{\rho\mu}g_{\sigma\lambda})\frac{k_\lambda}{2m_2}F_3(k^2)\Bigr\}D_{\mu\nu}(k)\times
\end{equation*}
\begin{equation*}
\left[g_{\rho\rho_1}-\frac{1}{2m_2^2}(m_2v_{\rho_1}-q_{\rho_1})(2m_2v_\rho-q_\rho)\right]
\left[g_{\sigma\sigma_1}-\frac{1}{2m_2^2}(m_2v_{\sigma_1}-p_{\sigma_1})(2m_2v_\sigma-p_\sigma)\right],
\end{equation*}
where the lepton vertex function $\Gamma_\mu=\frac{p_{1,\mu}+q_{1,\mu}}{2m_1}+(1+a_\mu)\sigma_{\mu\epsilon}
\frac{k_\epsilon}{2m_1}$, $a_\mu$ is the muon anomalous magnetic moment.
The Lorentz factors of the Dirac bispinors and transformed Lorentz factors of deuteron wave functions are written explicitly.
Inserting in \eqref{eq:q12} $\Psi_{\mu\nu}$ from \eqref{eq:q7}-\eqref{eq:q11}, averaging and summing over initial and
final state polarizations ${\cal M}$ and calculating the trace by means of the package Form \cite{form}
we find three matrix elements corresponding to $F=\frac{5}{2}$, $F=\frac{3}{2}$
$F=\frac{1}{2}$. The polarization sums for the fields with half-integer spin looks as follows \cite{fmms,berends,mosel}:
\begin{equation}
\label{eq:q13}
\hat\Pi_{\mu\nu}(F=3/2)=\sum_{{\cal M}=-F}^{F} \Psi^{\cal M}_{\mu}\bar\Psi^{\cal M}_{\nu}
=\frac{(\hat v+1)}{2}\left[g_{\mu\nu}-\frac{1}{3}\gamma_\mu\gamma_\nu-\frac{2}{3}v_\mu v_\nu+
\frac{1}{3}(v_\mu\gamma_\nu-v_\nu\gamma_\mu)\right],
\end{equation}
\begin{equation}
\label{eq:q14}
\hat\Pi_{\mu\nu;\rho\sigma}(F=5/2)=\sum_{{\cal M}=-F}^{F} \Psi^{\cal M}_{\mu\nu}\bar\Psi^{\cal M}_{\rho\sigma}=
\frac{(\hat v+1)}{2}\left[\frac{1}{2}\Bigl(P^1_{\mu\rho}P^1_{\nu\sigma}+P^1_{\mu\sigma}P^1_{\nu\rho}\right)-
\frac{1}{3}P^1_{\mu\nu}P^1_{\rho\sigma}-
\end{equation}
\begin{equation*}
-\frac{1}{10}\left(P^1_\mu P^1_\rho P^1_{\nu\sigma}+P^1_\nu P^1_\rho P^1_{\mu\sigma}+
P^1_\mu P^1_\sigma P^1_{\nu\rho}+P^1_\nu P^1_\sigma P^1_{\mu\rho}\right)
\Bigr],~~~P^1_{\mu\nu}=g_{\mu\nu}-v_\mu v_\nu,~~~P^1_\mu=P^1_{\mu\nu}\gamma_\mu.
\end{equation*}

Let us construct by this method basic hyperfine splittings of order $\alpha^4$. We project the amplitude \eqref{eq:q3} sequentially
on states with $j=1/2$, $F=1/2$ and $j=1/2$, $F=3/2$. Corresponding averaged amplitudes are the following:
\begin{equation}
\label{eq:qq1}
\overline{T_{1\gamma}({\bf p},{\bf q})}_{j=1/2}^{F=1/2}=\frac{\pi Z\alpha}{9} n_q^\delta n_p^\omega
Tr\Bigl\{(\hat v+1)(\gamma_{\rho_1}-v_{\rho_1})(\gamma_\delta+v_\delta)
\frac{[(m_1(\hat v+1)-{\boldsymbol\gamma}{\bf q}]}{2m_1}\times
\end{equation}
\begin{equation*}
\times\Gamma_\mu\frac{[m_1(\hat v+1)-{\boldsymbol\gamma}{\bf p}]}{2m_1}
(\gamma_\omega+v_\omega)(\gamma_{\sigma_1}-v_{\sigma_1})\Bigr\}D_{\mu\nu}(k)\times
\end{equation*}
\begin{equation*}
\Bigl\{g_{\rho\sigma}\frac{(p_2+q_2)_\nu}{2m_2}F_1(k^2)-\frac{(p_2+q_2)_\nu}{2m_2}\frac{k_\rho k_\sigma}{2m_2^2}F_2(k^2)+(g_{\rho\lambda}g_{\sigma\mu}-g_{\rho\mu}g_{\sigma\lambda})\frac{k_\lambda}{2m_2}F_3(k^2)\Bigr\}\times
\end{equation*}
\begin{equation*}
\times\left[g_{\rho\rho_1}-\frac{1}{2m_2^2}(m_2v_{\rho_1}-q_{\rho_1})(2m_2v_\rho-q_\rho)\right]
\left[g_{\sigma\sigma_1}-\frac{1}{2m_2^2}(m_2v_{\sigma_1}-p_{\sigma_1})(2m_2v_\sigma-p_\sigma)\right],
\end{equation*}
\begin{equation}
\label{eq:qq2}
\overline{T_{1\gamma}({\bf p},{\bf q})}_{j=1/2}^{F=3/2}=\frac{\pi Z\alpha}{6} n_q^\delta n_p^\omega
Tr\Bigl\{(\hat v+1)\hat\Pi_{\sigma_1\rho_1}(F=3/2)(\gamma_\delta-v_\delta)\gamma_5\times
\end{equation}
\begin{equation*}
\times\frac{[(m_1(\hat v+1)-{\boldsymbol\gamma}{\bf q}]}{2m_1}\Gamma_\mu\frac{[m_1(\hat v+1)-{\boldsymbol\gamma}{\bf p}]}{2m_1}\gamma_5
(\gamma_\omega-v_\omega)\Bigr\}D_{\mu\nu}(k)\times
\end{equation*}
\begin{equation*}
\Bigl\{g_{\rho\sigma}\frac{(p_2+q_2)_\nu}{2m_2}F_1(k^2)-\frac{(p_2+q_2)_\nu}{2m_2}\frac{k_\rho k_\sigma}{2m_2^2}F_2(k^2)+(g_{\rho\lambda}g_{\sigma\mu}-g_{\rho\mu}g_{\sigma\lambda})\frac{k_\lambda}{2m_2}F_3(k^2)\Bigr\}\times
\end{equation*}
\begin{equation*}
\times\left[g_{\rho\rho_1}-\frac{1}{2m_2^2}(m_2v_{\rho_1}-q_{\rho_1})(2m_2v_\rho-q_\rho)\right]
\left[g_{\sigma\sigma_1}-\frac{1}{2m_2^2}(m_2v_{\sigma_1}-p_{\sigma_1})(2m_2v_\sigma-p_\sigma)\right],
\end{equation*}
In the quasipotential method each of the amplitudes  \eqref{eq:qq1}-\eqref{eq:qq2} determines the interaction operator of
particles corresponding to states with selected quantum numbers. In this case, we get not only the contributions
of the hyperfine interaction, but also the Coulomb potential and a potential of fine structure. But the difference
\eqref{eq:qq1} and \eqref{eq:qq2} allows to find the hyperfine splitting of state $j=1/2$ which is written as an expression
of the output from the Form program:
\begin{equation}
\label{eq:qq3}
\overline{T_{1\gamma}({\bf p},{\bf q})}^{hfs}_{j=1/2}(F=3/2;1/2)=\frac{Z\alpha}{2}
\Biggl\{\frac{m_1}{m_2\kappa_d}\left[-\frac{pq}{{\bf k}^2}+
\frac{({\bf p}{\bf q})^2}{pq{\bf k}^2}\right]+
\end{equation}
\begin{equation*}
+(\kappa_d+1)\left[\frac{2({\bf p}{\bf q})^2}{pq{\bf k}^2}-\frac{(p^2+q^2)({\bf p}{\bf q})}{pq{\bf k}^2}\right]
+2(1+\kappa_d)(1+\frac{a_\mu}{2})\left[-\frac{pq}{{\bf k}^2}-
\frac{({\bf p}{\bf q})^2}{pq{\bf k}^2}+\frac{(p^2+q^2)({\bf p}{\bf q})}{pq{\bf k}^2}\right]
\Biggr\},
\end{equation*}
where $\kappa_d=0.714025 \mu_N$ is the deuteron anomalous magnetic moment \cite{codata2012}, connected
with the deuteron magnetic moment $\mu_d$ by the relation $\kappa_d=(\mu_d m_2/m_p-1)$.
In \eqref{eq:qq3} we take electromagnetic form factors at $k^2=0$ and
omit the quadrupole contribution which is studied in detail in next section. Normalization factor $3/4\pi$ coming from
wave function of orbital motion is taken into account.
Two other hyperfine splittings of $2P_{3/2}$ state looks as follows:
\begin{equation}
\label{eq:qq33}
\overline{T_{1\gamma}({\bf p},{\bf q})}^{hfs}_{j=3/2}(F=3/2;1/2)=
\frac{Z\alpha}{2}\Biggl\{\frac{m_1}{m_2\kappa_d}\left[-\frac{1}{2}\frac{pq}{{\bf k}^2}+\frac{11}{10}
\frac{({\bf p}{\bf q})^2}{pq{\bf k}^2}-\frac{3}{10}\frac{(p^2+q^2)({\bf p}{\bf q})}{pq{\bf k}^2}\right]+
\end{equation}
\begin{equation*}
+(\kappa_d+1)\left[\frac{({\bf p}{\bf q})^2}{pq{\bf k}^2}-\frac{1}{2}\frac{(p^2+q^2)({\bf p}{\bf q})}{pq{\bf k}^2}\right]
+\frac{2}{5}(1+\kappa_d)\left(1-\frac{a_\mu}{4}\right)\left[-\frac{pq}{{\bf k}^2}-
\frac{({\bf p}{\bf q})^2}{pq{\bf k}^2}+\frac{(p^2+q^2)({\bf p}{\bf q})}{pq{\bf k}^2}\right]-
\end{equation*}
\begin{equation*}
-\frac{({\bf p}{\bf q})}{pq}\left[\frac{3}{2}(1+\kappa_d)-\frac{3}{10}\frac{m_1\kappa_d}{m_2}-\frac{6}{5}
(1+\kappa_d)\left(1-\frac{a_\mu}{4}\right)\right]
\Biggr\},
\end{equation*}
\begin{equation}
\label{eq:qq55}
\overline{T_{1\gamma}({\bf p},{\bf q})}^{hfs}_{j=3/2}(F=5/2;3/2)=
\frac{Z\alpha}{2}\Biggl\{\frac{m_1}{m_2\kappa_d}\left[\frac{5}{6}\frac{pq}{{\bf k}^2}-\frac{1}{2}
\frac{({\bf p}{\bf q})^2}{pq{\bf k}^2}-\frac{1}{6}\frac{(p^2+q^2)({\bf p}{\bf q})}{pq{\bf k}^2}\right]+
\end{equation}
\begin{equation*}
+(\kappa_d+1)\left[-\frac{5}{3}\frac{({\bf p}{\bf q})^2}{pq{\bf k}^2}+\frac{5}{6}\frac{(p^2+q^2)({\bf p}{\bf q})}{pq{\bf k}^2}\right]
+\frac{2}{3}(1+\kappa_d)\left(1-\frac{a_\mu}{4}\right)\left[\frac{pq}{{\bf k}^2}+
\frac{({\bf p}{\bf q})^2}{pq{\bf k}^2}-\frac{(p^2+q^2)({\bf p}{\bf q})}{pq{\bf k}^2}\right]-
\end{equation*}
\begin{equation*}
-\frac{({\bf p}{\bf q})}{pq}\left[-\frac{5}{2}(1+\kappa_d)-\frac{1}{6}\frac{m_1\kappa_d}{m_2}+
2(1+\kappa_d)\left(1-\frac{a_\mu}{4}\right)\right]\Biggr\},
\end{equation*}
where the terms proportional to $({\bf p}{\bf q}/pq)$ vanish as a result of the angular integration.
They are important for the correct calculation of the vacuum polarization effects.
There are three types of integrals with the radial wave functions which are calculated analytically:
\begin{equation}
\label{eq:iii}
J_1=\int R_{21}(q)\frac{d{\bf q}}{(2\pi)^{3/2}}
\int R_{21}(p)\frac{d{\bf p}}{(2\pi)^{3/2}} \frac{pq}{({\bf p}-{\bf q})^2}=\Braket{\frac{pq}{({\bf p}-{\bf q})^2}}
=\frac{3}{16},
\end{equation}
\begin{equation*}
J_2=\Braket{\frac{({\bf p}{\bf q})^2}{pq({\bf p}-{\bf q})^2}}=\frac{5}{48},~~~J_3=\Braket{\frac{({\bf p}{\bf q})(p^2+q^2)}{pq({\bf p}-{\bf q})^2}}=\frac{5}{24}.
\end{equation*}
Note, that the terms on the right side of the equations~\eqref{eq:qq3}-\eqref{eq:qq55} proportional to $(1+\kappa_d)$
disappear after momentum integration and we obtain the following leading order contributions of diagonal matrix elements
to hyperfine splittings of $2P_{1/2}$ and $2P_{3/2}$ states:
\begin{equation}
\label{eq:hfs12}
\Delta E^{hfs}_{j=1/2}(F=3/2;1/2)=\frac{\alpha^4(1+\kappa_d)\mu^3}{12 m_1m_2}\left[1+
\frac{m_1\kappa_d}{2m_2(1+\kappa_d)}+\frac{a_\mu}{2}\right]=2070.5040~\mu eV,
\end{equation}
\begin{equation}
\label{eq:hfs321}
\Delta E^{hfs}_{j=3/2}(F=3/2;1/2)=\frac{\alpha^4(1+\kappa_d)\mu^3}{24 m_1m_2}
\left[\frac{2}{5}+\frac{m_1\kappa_d}{2m_2(1+\kappa_d)}-\frac{a_\mu}{10}\right]=420.9426~\mu eV,
\end{equation}
\begin{equation}
\label{eq:hfs322}
\Delta E^{hfs}_{j=3/2}(F=5/2;3/2)=\frac{5\alpha^4(1+\kappa_d)\mu^3}{72 m_1m_2}\left[\frac{2}{5}+\frac{m_1\kappa_d}{2m_2(1+\kappa_d)}-
\frac{a_\mu}{10}\right]=701.5712~\mu eV.
\end{equation}

Two amplitudes \eqref{eq:qq1}-\eqref{eq:qq2} are constructed combining firstly the orbital momentum and muon spin.
Then the spin of the nucleus is added. We can act slightly different expressing the states with $j=1/2$ and $j=3/2$ directly
in terms of introduced symmetrical and antisymmetrical states. This possibility is illustrated hereinafter.
In this method we can evaluate also off-diagonal matrix elements. Their calculation is demonstrated in next
section for quadrupole correction. To facilitate a comparison of the method of calculation and obtained contributions
to the previous approaches we make Appendix A, which demonstrates the calculation of corrections of order $\alpha^4$
in the coordinate representation. All basic contributions to hyperfine structure and numerous higher order corrections
are presented in Table~\ref{tb1}.

\begin{table}[htbp]
\caption{Diagonal matrix elements of hyperfine structure of $2P$-states in muonic deuterium}
\label{tb1}
\bigskip
\begin{tabular}{|c|c|c|c|c|c|}   \hline
The contribution   & $2^2P_{1/2}$, $\mu eV$ & $2^4P_{1/2}$, $\mu eV$& $2^2P_{3/2}$, $\mu eV$& $2^4P_{3/2}$, $\mu eV$& $2^6P_{3/2}$, $\mu eV$ \\   \hline
leading order $\alpha^4$      & -1380.3360   & 690.1680    & 8162.2889  & 8583.2315   & 9284.8027  \\
correction   &     &    &     &     &    \\  \hline
quadrupole correction    & 0    & 0  & 433.9033            & -347.1227     & 86.7807  \\
of order $\alpha^4$   &    &     &     &    &       \\   \hline
vacuum polarization    & -1.0706   & 0.5353    & -0.2802           & -0.1121  & 0.1681  \\
correction of order $\alpha^5$  &     &     &      &      &      \\    \hline
quadrupole and vacuum    & 0   & 0   & 0.3564            & -0.2851  & 0.0713   \\
polarization correction   &     &     &      &      &      \\
of order $\alpha^5$  &     &     &      &      &      \\    \hline
relativistic   & -0.1677   & 0.0838   & -0.0125    & -0.0050    & 0.0075   \\
correction of order $\alpha^6$ &    &     &      &     &         \\    \hline
vacuum polarization     & -0.0011                & 0.0005  & -0.0014     & -0.0006    & 0.0008      \\
correction of order $\alpha^6$&&&&&  \\    \hline
structure correction   & -0.0011               & 0.0021             & -0.0006            & 0.0010            & -0.0016     \\
of order $\alpha^6$   &                       &                     &                  &               &                     \\   \hline
Summary contribution        &  -1381.5765      & 690.7897   & 8596.2539     & 8235.7070   & 9371.8295 \\ \hline
\end{tabular}
\end{table}

\section{Quadrupole interaction corrections}

Quadrupole interaction originates from not completely spherical shape of the deuteron. If the potential of the muon
has also a non-spherical component at the position of the deuteron (only with muon angular momentum $j>1/2$) then there
exists a quadrupole energy shift \cite{sobel,drake,ibk}. Ordinary calculation of this contribution to hyperfine structure in muonic
deuterium is based on the representation of quadrupole interaction in coordinate space as a scalar product of two irreducible
tensor operators of rank 2. After that the matrix elements of tensor operators are expressed in terms of reduced matrix elements
using the Wigner-Eckart theorem.

In this work we develop another approach to the calculation of quadrupole interaction based on tensor
representation of P-wave projection operators describing muonic deuterium states.
In the case of $F=\frac{1}{2}$ and $F=\frac{3}{2}$ we should take the sum of two contributions regarding to
symmetric and antisymmetric wave projection function \eqref{eq:q8}-\eqref{eq:q11}. For completeness, we present two
averaged amplitudes corresponding to $\Psi_{\mu\nu}^S(F=\frac{3}{2})$ and $\Psi_{\mu\nu}^A(F=\frac{3}{2})$:
\begin{equation}
\label{eq:q15}
\overline{T_{1\gamma}({\bf p},{\bf q})}^A=\frac{\pi \alpha}{4} n_q^\delta n_p^\omega D_{\mu\nu}(k) Tr\Bigl\{
(\hat v+1)\hat\Pi_{\sigma\delta}(\gamma_\rho-v_\rho)\frac{[(m_1(1-\hat v)+{\boldsymbol\gamma}{\bf q}]}{2m_1}
\Gamma_\mu\times
\end{equation}
\begin{equation*}
\times\frac{[m_1(1-\hat v)+{\boldsymbol\gamma}{\bf p}]}{2m_1}(\gamma_\omega-v_\omega)+
(\rho\to\delta,\omega\to\sigma)-(\omega\to\sigma)-(\rho\to\delta)\Bigr\}\times
\end{equation*}
\begin{equation*}
\times\Bigl\{g_{\rho\sigma}\frac{(p_2+q_2)_\nu}{2m_2}F_1(k^2)-\frac{(p_2+q_2)_\nu}{2m_2}\frac{k_\rho k_\sigma}{2m_2^2}F_2(k^2)+(g_{\rho\lambda}g_{\sigma\mu}-g_{\rho\mu}g_{\sigma\lambda})\frac{k_\lambda}{2m_2}F_3(k^2)\Bigr\}=
\end{equation*}
\begin{equation*}
=-\frac{\pi \alpha Q_d}{3}\left[\frac{pq}{({\bf p}-{\bf q})^2}-\frac{({\bf p}{\bf q})\left(\frac{p}{q}+\frac{q}{p}\right)}{({\bf p}-{\bf q})^2}+
\frac{({\bf p}{\bf q})^2}{({\bf p}-{\bf q})^2}-\frac{1}{3}\frac{({\bf p}{\bf q})}{pq}\right],
\end{equation*}
\begin{equation}
\label{eq:q16}
\overline{T_{1\gamma}({\bf p},{\bf q})}^S=\frac{\pi \alpha}{20} n_q^\delta n_p^\omega D_{\mu\nu}(k)Tr\Bigl\{
(\hat v+1)\hat\Pi_{\sigma\delta}(\gamma_\rho-v_\rho)\frac{[(m_1(1-\hat v)+{\boldsymbol\gamma}{\bf q}]}{2m_1}
\Gamma_\mu\times
\end{equation}
\begin{equation*}
\times\frac{[m_1(1-\hat v)+{\boldsymbol\gamma}{\bf p}]}{2m_1}(\gamma_\omega-v_\omega)+
(\rho\to\delta,\omega\to\sigma)+(\omega\to\sigma)+(\rho\to\delta)\Bigr\}\times
\end{equation*}
\begin{equation*}
\times\Bigl\{g_{\rho\sigma}\frac{(p_2+q_2)_\nu}{2m_2}F_1(k^2)-\frac{(p_2+q_2)_\nu}{2m_2}\frac{k_\rho k_\sigma}{2m_2^2}F_2(k^2)+(g_{\rho\lambda}g_{\sigma\mu}-g_{\rho\mu}g_{\sigma\lambda})\frac{k_\lambda}{2m_2}F_3(k^2)\Bigr\}=
\end{equation*}
\begin{equation*}
=\frac{\pi \alpha Q_d}{15}\left[\frac{pq}{({\bf p}-{\bf q})^2}-\frac{({\bf p}{\bf q})\left(\frac{p}{q}+\frac{q}{p}\right)}{({\bf p}-{\bf q})^2}+
\frac{({\bf p}{\bf q})^2}{({\bf p}-{\bf q})^2}-\frac{1}{3}\frac{({\bf p}{\bf q})}{pq}\right],
\end{equation*}
where we keep only the contribution of the quadrupole form factor $F_Q(0)=Q_d$.
The index replacements designated in brackets of \eqref{eq:q15} and \eqref{eq:q16} refer to the written part of the amplitude.
Remaining integration with the radial wave functions is carried out analytically:
\begin{equation}
\label{eq:q17}
J=\int \frac{d{\bf p}}{(2\pi)^{3/2}} R_{12}(p)\int \frac{d{\bf q}}{(2\pi)^{3/2}} R_{12}(q)
\left[\frac{pq}{({\bf p}-{\bf q})^2}-\frac{({\bf p}{\bf q})\left(\frac{p}{q}+\frac{q}{p}\right)}{({\bf p}-{\bf q})^2}+
\frac{({\bf p}{\bf q})^2}{({\bf p}-{\bf q})^2}\right]=\frac{\mu^3(Z\alpha)^3}{16\pi}.
\end{equation}
The sum of \eqref{eq:q15} and \eqref{eq:q16} multiplied by the factor \eqref{eq:q17} gives the contribution to hyperfine
splitting $\frac{\alpha Q(\mu Z\alpha)^3}{48}\left(-\frac{4}{5}\delta_{F\frac{3}{2}}\right)$. Let us present final results for
other transitions:
\begin{equation}
\label{eq:q18}
\Delta E_Q^{hfs}=\frac{\alpha Q_d(\mu Z\alpha)^3}{48}\left[\delta_{F\frac{1}{2}}-\frac{4}{5}\delta_{F\frac{3}{2}}+\frac{1}{5}\delta_{F\frac{5}{2}}\right],
\end{equation}
The quadrupole moment of the deuteron is taken to be
$Q_d=0.285783(30)~fm^2$ \cite{quadr}. The result \eqref{eq:q18} coincides exactly with previous calculations made by different approaches \cite{borie3}.
As it follows from numerical values of \eqref{eq:q18} (see Table~\ref{tb1}) the quadrupole interaction changes the position
of levels $2^4P_{3/2}$ and $2^2P_{3/2}$.
Let us investigate in addition how the total angular momentum of the muon is changed in such transitions.
For this purpose, build again the amplitude of single-photon exchange combining consistently muon spin with the orbital angular momentum
and the deuteron spin. To be specific, we consider two diagonal matrix elements which are determined by averaged amplitudes
with $j=1/2$, $F=1/2$ and $j=1/2$, $F=3/2$:
\begin{equation}
\label{eq:q19}
\overline{T_{1\gamma}({\bf p},{\bf q})}_{\frac{1}{2}\frac{1}{2}}^{F=\frac{1}{2}}=\frac{\pi \alpha}{9} n_q^\delta n_p^\omega D_{\mu\nu}(k) Tr\Bigl\{
(\hat v+1)(\gamma_\rho-v_\rho)(\gamma_\delta+v_\delta)\times
\end{equation}
\begin{equation*}
\times\frac{[(m_1(1+\hat v)-{\boldsymbol\gamma}{\bf q}]}{2m_1}\Gamma_\mu
\frac{[m_1(1+\hat v)-{\boldsymbol\gamma}{\bf p}]}{2m_1}(\gamma_\omega+v_\omega)(\gamma_\sigma-v_\sigma)\Bigr\}\times
\end{equation*}
\begin{equation*}
\times\Bigl\{g_{\rho\sigma}\frac{(p_2+q_2)_\nu}{2m_2}F_1(k^2)-\frac{(p_2+q_2)_\nu}{2m_2}\frac{k_\rho k_\sigma}{2m_2^2}F_2(k^2)+(g_{\rho\lambda}g_{\sigma\mu}-g_{\rho\mu}g_{\sigma\lambda})\frac{k_\lambda}{2m_2}F_3(k^2)\Bigr\}=0,
\end{equation*}
\begin{equation}
\label{eq:q20}
\overline{T_{1\gamma}({\bf p},{\bf q})}_{\frac{1}{2}\frac{1}{2}}^{F=\frac{3}{2}}=\frac{\pi \alpha}{6} n_q^\delta n_p^\omega D_{\mu\nu}(k) Tr\Bigl\{
(\hat v+1)\hat\Pi_{\sigma\rho}(\gamma_\delta-v_\delta)\gamma_5\times
\end{equation}
\begin{equation*}
\times\frac{[(m_1(1+\hat v)-{\boldsymbol\gamma}{\bf q}]}{2m_1}\Gamma_\mu
\frac{[m_1(1+\hat v)-{\boldsymbol\gamma}{\bf p}]}{2m_1}\gamma_5(\gamma_\omega-v_\omega)\Bigr\}\times
\end{equation*}
\begin{equation*}
\times\Bigl\{g_{\rho\sigma}\frac{(p_2+q_2)_\nu}{2m_2}F_1(k^2)-\frac{(p_2+q_2)_\nu}{2m_2}\frac{k_\rho k_\sigma}{2m_2^2}F_2(k^2)+(g_{\rho\lambda}g_{\sigma\mu}-g_{\rho\mu}g_{\sigma\lambda})\frac{k_\lambda}{2m_2}F_3(k^2)\Bigr\}=0,
\end{equation*}
where lower indexes of the amplitude designate the muon total angular momentum.
With one side, the obtained expressions \eqref{eq:q19}-\eqref{eq:q20} explicitly show that quadrupole interaction does not contribute
to diagonal matrix elements with $j=1/2$. On the other side they demonstrate our choice of the tensor projectors on the
state with $j=1/2$:
\begin{equation}
\label{eq:q21}
\Psi_{\mu\nu}^{F=\frac{1}{2}}(j=1/2)=\frac{1}{3}\gamma_5(\gamma_\mu-v_\mu)\gamma_5(\gamma_\nu-v_\nu)\Psi,
\end{equation}
where the spinor $\Psi$ describes the state with total atomic momentum $F=\frac{1}{2}$. Using the Dirac algebra transformations
we can expand \eqref{eq:q21} on the basis $\Psi_{\mu\nu}^S(F=\frac{1}{2})$ and $\Psi_{\mu\nu}^A(F=\frac{1}{2})$:
\begin{equation}
\label{eq:q22}
\Psi_{\mu\nu}^{F=\frac{1}{2}}(j=1/2)=\frac{1}{\sqrt{3}}\Psi_{\mu\nu}^S(F=1/2)+
\sqrt{\frac{2}{3}}\Psi_{\mu\nu}^A(F=1/2).
\end{equation}
The same expansion can be performed for the state with $j=3/2$ and two states with $j=\frac{1}{2}$, $F=\frac{3}{2}$ and
$j=\frac{3}{2}$, $F=\frac{3}{2}$. They looks as follows:
\begin{equation}
\label{eq:q23}
\Psi_{\mu\nu}^{F=\frac{1}{2}}(j=3/2)=\sqrt{\frac{2}{3}}\Psi_{\mu\nu}^S(F=1/2)-
\sqrt{\frac{1}{3}}\Psi_{\mu\nu}^A(F=1/2),
\end{equation}
\begin{equation}
\label{eq:q24}
\Psi_{\mu\nu}^{F=\frac{3}{2}}(j=1/2)=\sqrt{\frac{5}{6}}\Psi_{\mu\nu}^S(F=3/2)-
\sqrt{\frac{1}{6}}\Psi_{\mu\nu}^A(F=3/2),
\end{equation}
\begin{equation}
\label{eq:q25}
\Psi_{\mu\nu}^{F=\frac{3}{2}}(j=3/2)=-\sqrt{\frac{1}{6}}\Psi_{\mu\nu}^S(F=3/2)+
\sqrt{\frac{5}{6}}\Psi_{\mu\nu}^A(F=3/2).
\end{equation}

Using \eqref{eq:q22}-\eqref{eq:q25} we can investigate off-diagonal matrix elements corresponding to different values of muon
angular momentum $j$. In fact, contributions with symmetric and antisymmetric tensor-spinor fields $\Psi_{\mu\nu}^S(F=\frac{1}{2},\frac{3}{2})$ and
$\Psi_{\mu\nu}^A(F=\frac{1}{2},\frac{3}{2})$ are evaluated above in matrix elements \eqref{eq:q15}-\eqref{eq:q16}.
Thus it is necessary to use only the correct coefficients of expansions \eqref{eq:q22}-\eqref{eq:q25}. As a result we obtain:
\begin{equation}
\label{eq:q26}
\Delta E^{hfs}_Q(j=1/2;j'=3/2)=\frac{\alpha Q_d(Z\mu\alpha)^3}{48}\left(\sqrt{2}\delta_{F\frac{1}{2}}-
\frac{1}{\sqrt{5}}\delta_{F\frac{3}{2}}\right).
\end{equation}
Numerically, all quadrupole corrections are large and presented in Table~\ref{tb1} and Table~\ref{tb2}.
Drawing attention to the significant value of the quadrupole corrections, we proceed to the consideration of other
important effects within the formulated framework.

\section{Vacuum polarization and structure corrections}

\begin{figure}[htbp]
\centering
\includegraphics[scale=0.8]{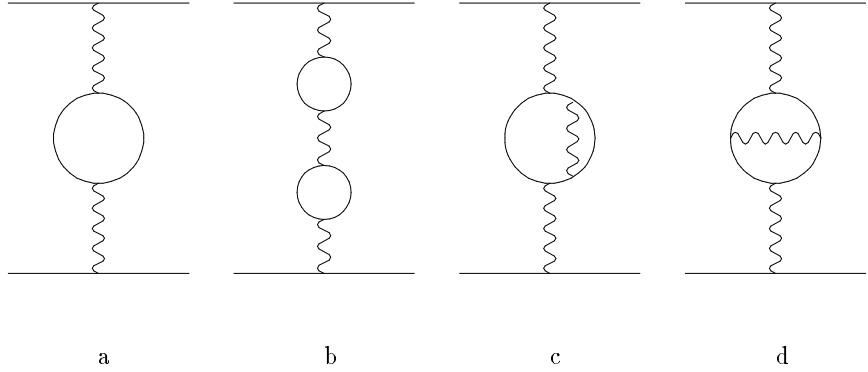}
\caption{Vacuum polarization effects in one-photon interaction. The wavy line represents hyperfine part of the interaction.}
\label{fig:hfsVP1}
\end{figure}

The above basic formulas for the amplitudes of the muon-deuteron interaction allow to calculate
the various corrections.
Next in importance are the corrections to the vacuum polarization (VP) of order $\alpha^5$. In the formulated framework
these effects can be easily studied. In the first order perturbation theory one-loop vacuum polarization
contribution to HFS is determined by the amplitude in Fig.~\ref{fig:hfsVP1}. For its calculation in momentum
representation which we use, the following replacement in the photon propagator should be done in \eqref{eq:qq3}:
\begin{equation}
\label{eq:replac}
\frac{1}{k^2}\to \frac{\alpha}{3\pi}\int_1^\infty\frac{\rho(\xi)d\xi}{k^2+4m_e^2\xi^2},~~~\rho(\xi)=\sqrt{\xi^2-1}(2\xi^2+1)/\xi^4.
\end{equation}
As a result we find that the vacuum polarization contribution to hyperfine splittings can be expressed in terms of three momentum
integrals which are a generalization of the three integrals discussed earlier in \eqref{eq:iii}:
\begin{equation}
\label{eq:jjj}
I_1=\int R_{21}(q)\frac{d{\bf q}}{(2\pi)^{3/2}}
\int R_{21}(p)\frac{d{\bf p}}{(2\pi)^{3/2}} \frac{pq}{({\bf p}-{\bf q})^2+4m_e^2\xi^2}=
\end{equation}
\begin{equation*}
=\Braket{\frac{pq}{({\bf p}-{\bf q})^2+4m_e^2\xi^2}}=\frac{a(3a+8)+6}{2(a+2)^4},~~~a=\frac{4m_e\xi}{\mu\alpha}.
\end{equation*}
\begin{equation*}
I_2=\Braket{\frac{({\bf p}{\bf q})^2}{pq({\bf p}-{\bf q})^2+4m_e^2\xi^2}}=\frac{a(3a+8)+10}{6(a+2)^4},~~~
I_3=\Braket{\frac{({\bf p}{\bf q})(p^2+q^2)}{pq({\bf p}-{\bf q})^2+4m_e^2\xi^2}}=\frac{2(4a+5)}{3(a+2)^4}.
\end{equation*}
Third integration over the spectral parameter $\xi$ also can be carried out analytically, but they are
quite cumbersome. So, we present here necessary VP correction to hyperfine splitting of $2P_{1/2}$ state
only in integral form:
\begin{equation}
\label{eq:evp}
\Delta E^{hfs}_{vp}(2P_{1/2})=\frac{\mu^3 \alpha(Z\alpha)^4}{3\pi m_1m_2}\int_1^\infty\rho(\xi)d\xi\Bigl[\frac{m_1\kappa_d}{2m_2}
\frac{(3a+2)}{3(a+2)^3}+(1+\kappa_d)\left(1+\frac{a_\mu}{2}\right)\frac{2(3a^2+4a+2)}{3(a+2)^4}-
\end{equation}
\begin{equation*}
-(1+\kappa_d)\frac{a^2}{2(a+2)^4}\Bigr]=1.0718~\mu eV.
\end{equation*}
The same calculation can be performed for the $2P_{3/2}$ state. The corresponding results are the following:
\begin{equation}
\label{eq:evp1}
\Delta E^{hfs}_{vp}(2P_{3/2})(F=3/2;1/2)=\frac{\mu^3 \alpha(Z\alpha)^4}{6\pi m_1m_2}\int_1^\infty\rho(\xi)d\xi\Bigl[\frac{m_1\kappa_d}{2m_2}
\frac{(3a+2)}{6(a+2)^3}+
\end{equation}
\begin{equation*}
+(1+\kappa_d)\left(1-\frac{a_\mu}{4}\right)\frac{(15a^2+8a+4)}{3(a+2)^4}
-(1+\kappa_d)\frac{2a^2}{(a+2)^4}\Bigr]=0.0595~\mu eV,
\end{equation*}
\begin{equation}
\label{eq:evp2}
\Delta E^{hfs}_{vp}(2P_{3/2})(F=5/2;3/2)=\frac{\mu^3 \alpha(Z\alpha)^4}{6\pi m_1m_2}\int_1^\infty\rho(\xi)d\xi\Bigl[\frac{m_1\kappa_d}{2m_2}
\frac{5(3a+2)}{18(a+2)^3}+
\end{equation}
\begin{equation*}
+(1+\kappa_d)\left(1-\frac{a_\mu}{4}\right)\frac{2(15a^2+8a+4)}{9(a+2)^4}
-(1+\kappa_d)\frac{10a^2}{3(a+2)^4}\Bigr]=0.0992~\mu eV.
\end{equation*}

Another important VP effect is related to quadrupole interaction discussed in previous section. Using for its
calculation basic expression \eqref{eq:qq3}, \eqref{eq:qq33}, \eqref{eq:qq55} and \eqref{eq:jjj} we obtain for diagonal
and off-diagonal matrix elements:
\begin{equation}
\label{eq:qvp}
\Delta E_{Q,vp}=\frac{\mu^3\alpha(Z\alpha)^4Q_d}{36\pi}\int_1^\infty\frac{(5a^2+8a+4)}{(a+2)^4}\rho(\xi)d\xi
\left[\delta_{F\frac{1}{2}}-\frac{4}{5}\delta_{F\frac{3}{2}}+\frac{1}{5}\delta_{F\frac{5}{2}}\right]=
\end{equation}
\begin{equation*}
=\left[\delta_{F\frac{1}{2}}-\frac{4}{5}\delta_{F\frac{3}{2}}+\frac{1}{5}\delta_{F\frac{5}{2}}\right]\times 0.2441~\mu eV,
\end{equation*}
\begin{equation}
\label{eq:qvp1}
\Delta E_{Q,vp}(j=3/2;j'=1/2)=\frac{\mu^3\alpha(Z\alpha)^4Q_d}{72\pi}\int_1^\infty\frac{(5a^2+24a+24)}{3(a+2)^4}\rho(\xi)d\xi
\left[\sqrt{2}\delta_{F\frac{1}{2}}-\frac{1}{\sqrt{5}}\delta_{F\frac{3}{2}}\right]=
\end{equation}
\begin{equation*}
=\left[\sqrt{2}\delta_{F\frac{1}{2}}-\frac{1}{\sqrt{5}}\delta_{F\frac{3}{2}}\right]\times 0.0630~\mu eV.
\end{equation*}

A comparison of our results \eqref{eq:evp}-\eqref{eq:qvp1} with earlier estimates in \cite{borie3}
shows that there is a significant difference of the order of tenths of $\mu eV$.
For this reason we decided to perform additional validation of our results using a different method of the calculation.
As was shown in \cite{apm2008} the vacuum polarization effects presented in Fig.~\ref{fig:hfsVP1} in first order perturbation
theory, can be calculated in coordinate representation. The amplitude shown in Fig.~\ref{fig:hfsVP1}(a) gives the following
hyperfine interaction potential in coordinate space:
\begin{gather}
\notag\Delta V^{hfs}_{1\gamma,vp}(r)=\frac{Z\alpha(1+\kappa_d)}{2m_1 m_2 r^3} \frac{\alpha}{3\pi}\int_1^\infty\rho(\xi)d\xi e^{-2m_e\xi r} \biggl\{\biggl( 1+\frac{m_1\kappa_d}{m_2 (1+\kappa_d)} \biggl)\times\\
\label{eq:1gammavppot}
\times(\boldsymbol L\cdot\boldsymbol s_2)
(1+2m_e\xi r)-(1+a_\mu)\biggl(4m_e^2 \xi^2 r^2[(\boldsymbol s_1\cdot \boldsymbol s_2)-(\boldsymbol s_1\cdot\boldsymbol n)(\boldsymbol s_2\cdot \boldsymbol n)]+\\
\notag+(1+2 m_e \xi r)[(\boldsymbol s_1\cdot \boldsymbol s_2)-3(\boldsymbol s_1\cdot\boldsymbol n)(\boldsymbol s_2\cdot \boldsymbol n)]\biggl)\biggl\}.
\end{gather}
Averaging \eqref{eq:1gammavppot} over the Coulomb wave functions, we obtain an analytical expression for the vacuum polarization
correction of order $\alpha^5$ in one-photon interaction:
\begin{gather}
\notag \Delta E^{hfs}_{1\gamma,vp}(r)=\frac{\alpha^4 \mu^3(1+\kappa_d)}{24m_1 m_2 r^3} \frac{\alpha}{6\pi}\int_1^\infty\rho(\xi)d\xi \int_0^\infty x dx e^{-x[1+\frac{2m_e\xi}{W}]} \biggl[\biggl( 1+\frac{m_1\kappa_d}{m_2 (1+\kappa_d)} \biggl)\times\\
\label{eq:1gammavp}
\times\overline{T_1}
(1+\frac{2m_nj;e\xi}{W} x)-(1+a_\mu)\biggl(\frac{4m_e^2 \xi^2 x^2}{W^2}\overline{T_3}+(1+\frac{2 m_e \xi}{W} x)\overline{T_2}\biggl)\biggl],
\end{gather}
where we introduce the designations for operators $T_i$ in \eqref{eq:1gammavppot}:
\begin{equation}
\label{eq:t123}
T_1=(\boldsymbol L\cdot\boldsymbol s_2),~~~
T_2=\biggl[(\boldsymbol s_1\cdot \boldsymbol s_2)-3(\boldsymbol s_1\cdot\boldsymbol n)(\boldsymbol s_2\cdot \boldsymbol n)\biggl],~~~
T_3=\biggl[(\boldsymbol s_1\cdot \boldsymbol s_2)-(\boldsymbol s_1\cdot\boldsymbol n)(\boldsymbol s_2\cdot \boldsymbol n)\biggl].
\end{equation}

The coordinate integration in \eqref{eq:1gammavp} is carried out analytically and numerically over the spectral parameter $\xi$.
Numerical results for separate states include both diagonal and off-diagonal matrix elements:
\begin{equation}
\Delta E^{hfs}_{j=1/2,vp}(F=1/2)=-0.7145~\mu eV,
\end{equation}
\begin{equation*}
\Delta E^{hfs}_{j=1/2,vp}(F=3/2)=0.3573~\mu eV,
\end{equation*}
\begin{equation*}
\Delta E^{hfs}_{j=3/2,vp}(F=1/2)=-0.0992~\mu eV,
\end{equation*}
\begin{equation*}
\Delta E^{hfs}_{j=3/2,vp}(F=3/2)=-0.0397~\mu eV,
\end{equation*}
\begin{equation*}
\Delta E^{hfs}_{j=3/2,vp}(F=5/2)= 0.0595~\mu eV.
\end{equation*}
\begin{equation*}
\Delta E^{hfs}_{(j=1/2 \to j=3/2),vp}(F=1/2)=-0.1111~\mu eV,
\end{equation*}
\begin{equation*}
\Delta E^{hfs}_{(j=1/2 \to j=3/2),vp}(F=3/2)= -0.1757~\mu eV.
\end{equation*}
They evidently show that two our approaches to the calculation of hyperfine structure in muonic deuterium P-states lead to the same results.
Two-loop vacuum polarization corrections shown in Fig.~\ref{fig:hfsVP1} are calculated in a similar way. They are included in Appendix C.
Their numerical value is essentially smaller (see Table~\ref{tb1}).

\begin{figure}[htp]
\centering
\includegraphics[scale=0.8]{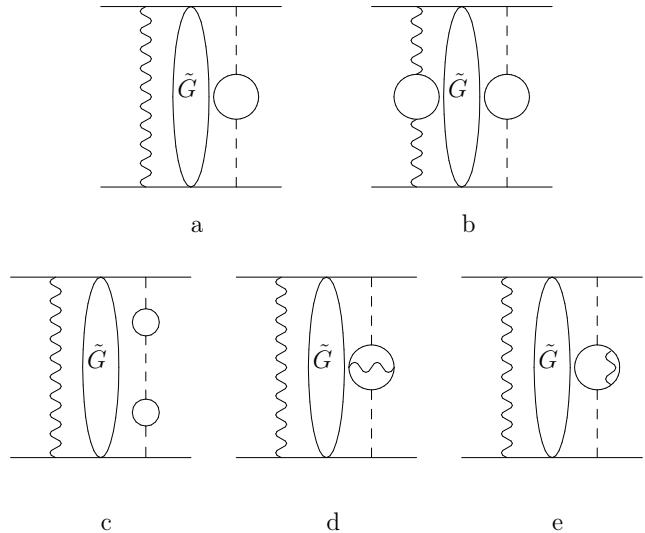}
\caption{Vacuum polarization effects in the second order perturbation theory. Dashed and wavy lines represent correspondingly the Coulomb
and hyperfine interactions.}
\label{fig:hfsVPSOPT}
\end{figure}

For a completeness, we analyze vacuum polarization corrections of order $\alpha^5$ in second order
perturbation theory, which are determined by the reduced Coulomb Green's function \cite{hameka,pachucki}
(see the amplitude in Fig.~\ref{fig:hfsVPSOPT}(a)):
\begin{equation}
\label{eq:green}
G_{2P}(\boldsymbol r,\boldsymbol r')=-\frac{\mu^2(Z\alpha)}{36z^2z'^2}\biggl( \frac{3}{4\pi}
\boldsymbol n \boldsymbol n'\biggl) e^{-(z+z')/2}g(z,z'),
\end{equation}
\begin{gather*}
g(z,z')=24z_<^3+36z_<^3z_>+36z_<^3z_>^2+24z_>^3+36z_<z_>^3+36z_<^2z_>^3+49z_<^3z_>^3-3z_<^4z_>^3-\\
-12e^{z_<}(2+z_<+z_<^2)z_>^3-3z_<^3z_>^4+12z_<^3z_>^3[-2C+Ei(z_<)-ln z_<-ln z_>],
\end{gather*}
where $C=0.5772...$ is the Euler constant, $z=Wr,~z_<=min(z,z'),~z_>=max(z,z')$.
Using \eqref{eq:eq1a} and \eqref{eq:green}, we obtain the following integral expression for VP correction \cite{apm2008}:
\begin{gather}
\notag \Delta E_{vp,SOPT}^{hfs}=\frac{\alpha^4\mu^3(1+\kappa_d)}{24m_1 m_2}\frac{\alpha}{54\pi}\int_1^\infty\rho(\xi)d\xi\int_0^\infty dx \int_0^\infty\frac{e^{-x'}}{x'^2}dx'e^{-x\biggl(1+\frac{2m_e\xi}{W} \biggl)}\times \\
\label{eq:SOPTres}
\times\biggl[ \overline{T_1}+\frac{m_1 \kappa_{d}}{m_2(1+\kappa_{d})}\overline{T_1}-(1+a_{\mu})\overline{T_2} \biggl].
\end{gather}
Similarly, the correction of vacuum polarization and quadrupole interaction in second order PT has the form:
\begin{equation}
\label{eq:SOPTquadr}
\Delta E_{vp,Q,SOPT}^{hfs}=\frac{\alpha^5\mu^3 Q_d}{2592\pi}\int_1^\infty\rho(\xi)d\xi\int_0^\infty dx \int_0^\infty
\frac{e^{-x'}}{x'^2}dx'e^{-x\biggl(1+\frac{2m_e\xi}{W} \biggl)}g(x,x')\times
\end{equation}
\begin{equation*}
\times\overline{\biggl[(\boldsymbol s_2\cdot \boldsymbol s_2)-3(\boldsymbol s_2\cdot\boldsymbol n)(\boldsymbol s_2\cdot \boldsymbol n)\biggl]}
=\Biggl\{{
{\left( \delta_{F,\frac{1}{2}}-\frac{4}{5} \delta_{F,\frac{3}{2}}+\frac{1}{5}\delta_{F,\frac{5}{2}}\right) \cdot 0.1120~(\mu eV),~~~j=j'=\frac{3}{2}}\atop
{\left( \sqrt{2}\delta_{F,\frac{1}{2}}-\frac{1}{\sqrt{5}} \delta_{F,\frac{3}{2}}\right) \cdot 0.1120~(\mu eV),~~~j=\frac{3}{2}, j'=\frac{1}{2}.}}
\end{equation*}

The coordinate integration over $x,~x'$ is performed again analytically and numerically over $\xi$. Summary numerical values of
contributions in the first and second orders of PT to the P-state energies are presented in Table~\ref{tb1},
Table~\ref{tb2} separately for diagonal and off-diagonal matrix elements.

Based on the amplitudes \eqref{eq:qq3}-\eqref{eq:qq55} it is possible to find the nuclear structure correction (the index str designates
this contribution) to hyperfine
splittings. For this aim, we will introduce into them an additional factor $(-r_d^2{\bf k}^2/6)$ with the deuteron root mean square radius
connected with the expansion of form factors and omit factors containing the deuteron magnetic moment.
After evident simplifications we obtain the following contributions to hyperfine
splitting potentials for states $2P_{1/2}$ and $2P_{3/2}$:
\begin{equation}
\label{eq:pq3}
\overline{T_{1\gamma,str}({\bf p},{\bf q})}^{hfs}_{j=1/2}(F=3/2;1/2)=\frac{Z\alpha r_d^2}{12}
\bigl\{\frac{m_1}{m_2}\bigl[pq-\frac{({\bf p}{\bf q})^2}{pq}\bigr]-2\frac{({\bf p}{\bf q})^2}{pq}
+2(1+\frac{a_\mu}{2})\bigl[pq+\frac{({\bf p}{\bf q})^2}{pq}\bigr]\bigr\},
\end{equation}
\begin{equation}
\label{eq:pq33}
\overline{T_{1\gamma,str}({\bf p},{\bf q})}^{hfs}_{j=3/2}(F=3/2;1/2)=\frac{Z\alpha r_d^2}{12}
\bigl\{\frac{m_1}{2m_2}\bigl[pq-\frac{({\bf p}{\bf q})^2}{pq}\bigr]-4\frac{({\bf p}{\bf q})^2}{pq}
+\frac{2}{5}(1-\frac{a_\mu}{4})\bigl[pq+7\frac{({\bf p}{\bf q})^2}{pq}\bigr]\bigr\},
\end{equation}
\begin{equation}
\label{eq:pq55}
\overline{T_{1\gamma,str}({\bf p},{\bf q})}^{hfs}_{j=3/2}(F=5/2;3/2)=\frac{Z\alpha r_d^2}{12}
\bigl\{\frac{5m_1}{6m_2}\bigl[\frac{({\bf p}{\bf q})^2}{pq}-pq\bigr]+\frac{20}{3}\frac{({\bf p}{\bf q})^2}{pq}
-\frac{2}{3}(1-\frac{a_\mu}{4})\bigl[pq+7\frac{({\bf p}{\bf q})^2}{pq}\bigr]\bigr\}.
\end{equation}
Further integration and consideration of the general normalization factor directly lead to the following splittings:
\begin{equation}
\label{eq:sp1}
\Delta E^{hfs}_{1\gamma,str}(j=1/2,F=3/2;1/2)=\frac{\mu^5\alpha^6 r_d^2}{16m_1m_2}
\left(\frac{m_1}{m_2}+\frac{a_\mu}{2}\right)=0.0032~\mu eV,
\end{equation}
\begin{equation}
\label{eq:sp2}
\Delta E^{hfs}_{1\gamma,str}(j=3/2,F=3/2;1/2)=\frac{\mu^5\alpha^6 r_d^2}{32m_1m_2}
\left(\frac{m_1}{m_2}-a_\mu\right)=0.0016~\mu eV,
\end{equation}
\begin{equation}
\label{eq:sp3}
\Delta E^{hfs}_{1\gamma,str}(j=3/2,F=5/2;3/2)=-\frac{5\mu^5\alpha^6 r_d^2}{96m_1m_2}
\left(\frac{m_1}{m_2}-a_\mu\right)=-0.0026~\mu eV.
\end{equation}

As expected, these corrections are very small and do not affect the comparison of theoretical results and
planned experimental data. Other corrections of order $\alpha^6$ are discussed in two Appendixes B and C.

\begin{table}[htbp]
\caption{Off-diagonal matrix elements in the hyperfine structure of P-wave muonic deuterium.}
\label{tb2}
\bigskip
\begin{tabular}{|c|c|c|}   \hline
Contribution to HFS       & $2^2P_{1/2,3/2}$, ìêýÂ& $2^4P_{1/2,3/2}$, ìêýÂ \\   \hline
leading order $\alpha^4$        & -126.0372           & -199.2824          \\
correction   &      &    \\   \hline
quadrupole correction    &  613.6320   &  -194.0475     \\
of order $\alpha^4$ &    &    \\   \hline
vacuum polarization      &  -0.1437  &   -0.2271     \\
correction of order $\alpha^5$  &   &  \\   \hline
quadrupole and vacuum    & 0.0891  & -0.0282    \\
polarization correction   &     &       \\
of order $\alpha^5$  &     &        \\    \hline
relativistic correction    & -0.0043              & -0.0067              \\
of order $\alpha^6$ &    &    \\  \hline
vacuum polarization      & 0.0001 &  0.0001  \\
correction of order $\alpha^6$  &    &    \\    \hline
Summary contribution  &  487.5360 &  -393.5918\\   \hline
\end{tabular}
\end{table}

\section{Summary and discussion}

In this work we investigate the hyperfine structure of energy levels related to the P-wave states of muonic deuterium
on the basis of three dimensional quasipotential approach in quantum electrodynamics. To increase the accuracy
of the calculation we take into account the leading order contribution and several basic corrections of order $\alpha^5$
and $\alpha^6$. These corrections are connected with the vacuum polarization effect, quadrupole interaction, nuclear
structure and relativistic effects. Some corrections are obtained in analytical form, but most part
of contributions to the energy spectrum
is presented first in integral form, and then calculated numerically. All results are presented in
Tables~\ref{tb1},\ref{tb2},\ref{tb3} giving the values of diagonal and off-diagonal matrix elements and the positions of the P-energy levels.

\begin{table}[htbp]
\caption{Hyperfine structure of P-states in muonic deuterium}
\label{tb3}
\bigskip
\begin{tabular}{|c|c|c|}   \hline
State       & Energy, meV , \cite{borie3},& Energy, meV\\ \hline
$2^2P_{1/2}$        & -1.4056  & -1.40534\\ \hline
$2^4P_{1/2}$        & 0.6703 & 0.67031\\ \hline
$2^2P_{3/2}$        & 8.6194 &   8.62002 \\ \hline
$2^4P_{3/2}$        & 8.2560 & 8.25618\\ \hline
$2^6P_{3/2}$        & 9.3729  & 9.37183 \\ \hline
\end{tabular}
\end{table}

We would like to point out three main our results obtained in this work.

1. New approach based on the use of special type projection operators on the states with definite quantum
numbers of atomic angular momentum $F$ and total muon angular momentum $j$ is developed. It allows to simplify
essentially the construction of the particle interaction operator through the use of computer methods
for calculating Feynman amplitudes. In particular, this method can be useful when working with different loop corrections.

2. We have increased the accuracy of the calculation of P-wave hyperfine splittings primarily due to the correct
account the corrections of the fifth order over $\alpha$. To this end, the contributions have been built into the
operator of the interaction of particles that are connected to the vacuum polarization and quadrupole interactions.
We check the obtained results in two ways, in the formulated framework of tensor projection operators in momentum representation
and the more traditional method for the calculation of corrections in the energy spectrum in the coordinate representation.
Moreover, in our calculation we take into account the contributions not only the first but also the second-order
perturbation theory.

3. New higher order $O(\alpha^6)$ corrections are calculated. These corrections although small numerically and do not
affect on the comparison with future experimental data, but clarify the structure of the perturbation series
for the hyperfine splittings.

Let us present more detail comparison of the results with previous calculations in \cite{borie3,brodsky}.
Being different in the method of obtaining corrections of leading order $O(\alpha^4)$ our results coincide with
\cite{borie3,brodsky}.
We mean both the spin-orbit, spin-spin contributions of order $O(\alpha^4)$ \cite{borie3,brodsky} and quadrupole corrections
of the same order \cite{borie3}.
But we obtain the fifth order in $\alpha$ corrections which are differ significantly from the results of \cite{borie3}.
In paper \cite{borie3} the vacuum polarization corrections to hyperfine part of the Breit Hamiltonian are determined by
the following modification of the potential with $l>0$:
\begin{equation}
\label{eq:borie_p}
\frac{1}{r}\frac{dV}{dr}=\frac{Z\alpha}{r^3}\left[1+\frac{\alpha}{3\pi}\int_1^\infty\rho(\xi)d\xi(1+2m_e\xi r)e^{-2m_e\xi r}\right].
\end{equation}
This leads to appearing of special factor of the form $(1+\varepsilon_{2P})$ with numerical value $\varepsilon_{2P}=0.000391$
for the quadrupole correction and
\begin{equation}
\label{eq:comp}
\varepsilon_{2P}=\frac{\alpha}{3\pi}\int_1^\infty\rho(\xi)d\xi\left(\frac{1}{(1+a\xi)^2}+\frac{2az}{(1+a\xi)^3}\right)
\end{equation}
for the Uehling correction to the Breit Hamiltonian. Numerically, the coefficient in \eqref{eq:comp} is equal
to the same value $\varepsilon_{2P}=0.000391$. In our calculation we demonstrate that the vacuum polarization
corrections to P-states are determined by different potentials (compare \eqref{eq:borie_p} with our formula \eqref{eq:1gammavppot})
and have different form for states with various quantum numbers
$F$ and $j$. In contrast to \cite{borie3} we have performed exact construction of corresponding potentials for
different P-states and obtained through them numerical results that can not be reduced to a factor \eqref{eq:comp}.
Our results are checked by two independent methods. As an example, we give a comparison of our vacuum polarization plus quadrupole interaction
contributions to hyperfine splittings of the level $2P_{3/2}$ with the results of \cite{borie3}. In \cite{borie3} these contributions are
equal to $\Delta \tilde E^{hfs}_{j=3/2}(F=3/2;1/2)=-3\mu^3\alpha^4 Q_d\varepsilon_{2p}/80=-0.3058~\mu eV$,
$\Delta \tilde E^{hfs}_{j=3/2}(F=5/2;3/2)=\mu^3\alpha^4 Q_d\varepsilon_{2p}/48=0.1699~\mu eV$ and differ essentially from our
corresponding values $(-0.4394)~\mu eV$ and $0.2441~\mu eV$. The same situation occurs for other VP corrections.

Summing all diagonal and off-diagonal matrix elements we obtain the following energy matrix
\begin{widetext}
\begin{equation}
\label{eq:2P_matrix}
\small{
\begin{aligned}
M ~
  &= & & \bordermatrix{
              &2^2P_{1/2}&2^4P_{1/2}&2^2P_{3/2}&2^4P_{3/2}&2^6P_{3/2}\\[1ex]
2^2P_{1/2} & -1381.5765 &  0          & 487.5360   &  0          &  0          \\[1ex]
2^4P_{1/2} &  0          &  690.7897&  0           & -393.5918   &  0          \\[1ex]
2^2P_{3/2} &  487.5360  &  0          &  8596.2539 &  0          &  0          \\[1ex]
2^4P_{3/2} &  0          & -393.5918 &  0           &  8235.7070&  0          \\[1ex]
2^6P_{3/2} &  0          &  0          &  0           &  0          &  9371.8295 \\[1ex]
} ~ \mathrm{\mu eV.}
\end{aligned}
}
\end{equation}
\end{widetext}

Its diagonalization leads directly to the position of the energy levels $2P$ (see Table~\ref{tb3}) and hyperfine splitting intervals
which can be measured in the experiment. Accounting the accuracy of the calculation, we have added one extra decimal place in our results
in Table~\ref{tb3}.

\begin{acknowledgments}
We are grateful to E.~Borie, F.~Kottmann and R.~Pohl for valuable information about CREMA experiments,
critical remarks and useful discussion of different questions related to the energy levels of light muonic atoms.
The work is supported by the Russian Foundation for Basic Research (grant 14-02-00173), the Ministry of Education and
Science of Russia under Competitiveness Enhancement Program 2013-2020 and the Dynasty Foundation.
\end{acknowledgments}

\appendix

\section{Basic contributions to hyperfine structure in coordinate representation}
\label{app:fis}

Basic
contribution to hyperfine structure is determined by hyperfine part of the Breit Hamiltonian \cite{BS}:
\begin{equation}
\label{eq:eq1a}
\Delta V^{hfs}_B(r)=\frac{Z\alpha (1+\kappa_{d})}{2m_1 m_2 r^3}\bigl[1+\frac{m_1\kappa_{d}}{m_2(1+\kappa_{d})}
\bigr](\boldsymbol L\boldsymbol s_2)
-\frac{Z\alpha (1+\kappa_{d})(1+a_{\mu})}{2m_1 m_2 r^3}\left[(\boldsymbol s_1\boldsymbol s_2)-
3(\boldsymbol s_1\boldsymbol n)(\boldsymbol s_2\boldsymbol n)\right],
\end{equation}
where $m_1$, $m_2$ are the muon and deuteron masses, $\kappa_d$, $a_{\mu}$ are the deuteron and muon anomalous
magnetic moments, ${\boldsymbol s}_1$ and ${\boldsymbol s}_2$ are the spin operators of muon and deuteron,
$\boldsymbol n = {\boldsymbol r}/r.$ The operator \eqref{eq:eq1a} does not commute with the muon total angular momentum
${\bf J}={\bf L}+{\bf s}_1$. As a result there is the mixing between energy levels $2P_{1/2}$ and $2P_{3/2}$.

For the calculation of diagonal matrix elements $\Braket{2P_{1/2}|\Delta V^{hfs}_B|2P_{1/2}}$ and
$\Braket{2P_{3/2}|\Delta V^{hfs}_B|2P_{3/2}}$ we use the Coulomb wave function of $2P$-state in coordinate representation:
\begin{equation}
\label{eq:eq2a}
\Psi_{2P}(\boldsymbol r)=\frac{1}{2\sqrt{6}}W^{\frac{5}{2}}re^{-\frac{Wr}{2}}Y_{1m}(\theta,\phi), W=\mu Z \alpha.
\end{equation}
The angle averaging in \eqref{eq:eq1a} can be carried out by means of the following replacements \cite{BS}:
\begin{equation}
\label{eq:eq3a}
{\boldsymbol s}_1 \rightarrow \boldsymbol J \frac{\overline{({\boldsymbol s}_1\cdot {\boldsymbol J})}}{J^2},
{\boldsymbol L} \rightarrow \boldsymbol J \frac{\overline{({\boldsymbol L} \cdot{\boldsymbol J})}}{J^2},
\end{equation}
which give the eigenvalues of the corresponding operators:
\begin{equation}
\label{eq:eq4a}
\overline{({\boldsymbol s}_1 \cdot{\boldsymbol J})}=\frac{1}{2}\biggl[j(j+1)-l(l+1)+\frac{3}{4} \biggl],
\overline{({\boldsymbol L}\cdot {\boldsymbol J})}=\frac{1}{2}\biggl[j(j+1)+l(l+1)-\frac{3}{4} \biggl],
\end{equation}
\begin{equation}
\label{eq:eq5a}
\Braket{\delta_ij-3n_i n_j}=-\frac{1}{5}(4\delta_{ij}-3L_i L_j-3L_j L_i).
\end{equation}
The diagonal matrix elements have the general form:
\begin{equation}
\label{eq:eq7a}
E^{hfs}_B=\frac{\alpha^4\mu^3(1+\kappa_{d})}{48m_1 m_2}\biggl[ \overline{T_1}+
\frac{m_1 \kappa_{d}}{m_2(1+\kappa_{d})}\overline{T_1}-(1+a_{\mu})\overline{T_2} \biggl],
\end{equation}
where the operators $T_i$ are defined in \eqref{eq:t123}.
Substituting here $\overline{T_1}$ and $\overline{T_2}$ for definite quantum numbers $F$ and $j$, we obtain the leading
order contributions to the hyperfine structure of $2P_{1/2}$ and $2P_{3/2}$ states:
\begin{equation}
\label{eq:eq8a}
^{2}E^{hfs}_{1/2}=-\frac{\alpha^4\mu^3(1+\kappa_{d})}{18m_1 m_2}\biggl[ 1+
\frac{m_1 \kappa_{d}}{2m_2(1+\kappa_{d})}+\frac{a_{\mu}}{2}\biggl]=-1380.3360~\mu eV,
\end{equation}
\begin{equation}
\label{eq:eq9a}
^{4}E^{hfs}_{1/2}=\frac{\alpha^4\mu^3(1+\kappa_{d})}{36m_1 m_2}\biggl[ 1+
\frac{m_1 \kappa_{d}}{2m_2(1+\kappa_{d})}+\frac{a_{\mu}}{2}\biggl]=690.1680~\mu eV,
\end{equation}
\begin{equation}
\label{eq:eq10a}
^{2}E^{hfs}_{3/2}=-\frac{\alpha^4\mu^3(1+\kappa_{d})}{72m_1 m_2}\biggl[2+
\frac{5m_1 \kappa_{d}}{2m_2(1+\kappa_{d})}-\frac{a_{\mu}}{2}\biggl]=8162.2889~\mu eV,
\end{equation}
\begin{equation}
\label{eq:eq11a}
^{4}E^{hfs}_{3/2}=-\frac{\alpha^4\mu^3(1+\kappa_{d})}{36m_1 m_2}\biggl[\frac{2}{5}+
\frac{m_1 \kappa_{d}}{2m_2(1+\kappa_{d})}-\frac{a_{\mu}}{10}\biggl]=8583.2315~\mu eV,
\end{equation}
\begin{equation}
\label{eq:eq12a}
^{6}E^{hfs}_{3/2}=-\frac{\alpha^4\mu^3(1+\kappa_{d})}{24m_1 m_2}\biggl[\frac{2}{5}+
\frac{m_1 \kappa_{d}}{2m_2(1+\kappa_{d})}-\frac{a_{\mu}}{10}\biggl]=9284.8027~\mu eV,
\end{equation}
where we take into account the fine structure interval $\Delta E_{fs}=8.86386~meV$ calculated in \cite{borie3,mk2011}.
All expressions \eqref{eq:eq8a}-\eqref{eq:eq12a} contain the correction to the anomalous magnetic moment of the muon.

Off-diagonal matrix elements
$\Braket{2P_{1/2}|\Delta V^{hfs}|2P_{3/2}}^{F=1/2}$ and $\Braket{2P_{1/2}|\Delta V^{hfs}|2P_{3/2}}^{F=3/2}$ are
essential to achieve a high accuracy of the calculation.
They differ by the value of atomic angular momentum. The angular averaging by means of \eqref{eq:eq5a} leads to
$\overline T_1=2\overline T_2$. For the calculation $\overline{({\bf L}{\bf s}_2)}$, we use the general formula
for the matrix elements of the scalar product of two irreducible tensor operators:
\begin{equation}
\label{eq:t1t2fin}
\Braket{j' s_2 F|(T^1\cdot T^2)|j s_2 F}=(-1)^{s_2+J'-F}W(j s_2 j' s_2;F 1)\Braket{j'||T^1||j}\Braket{s_2||T^2||s_2},
\end{equation}
where $W(j s_2 j' s_2;F 1)$ is the Racah coefficient. Applying \eqref{eq:t1t2fin} to $\overline{({\bf L}{\bf s}_2)}$
we find:
\begin{gather}
\Braket{j' s_2 F|({\boldsymbol L}\cdot{\boldsymbol s}_2)|j s_2 F}=(-1)^{-j-F-s_2+L+3/2+j'}\sqrt{(2j'+1)(2j+1)}\times\\
\label{eq:t1t2off1}
\times\sqrt{(2s_2+1)(s_2+1)s_2(2L+1)(L+1)L}\biggl\{
\begin{array}{ccc}
j & s_2 & F \\
s_2 & j' & 1
\end{array}
\biggl\}\biggl\{
\begin{array}{ccc}
l & j' & \frac{1}{2} \\
j & l & 1
\end{array}
\biggl\}.
\end{gather}
Two off-diagonal matrix elements of the operator $T_1$ have the form:
\begin{equation}
\label{eq:offdavg}
\Braket{\frac{1}{2},1,\frac{1}{2}|(\boldsymbol L\cdot\boldsymbol s_2)|\frac{3}{2},1,\frac{1}{2}}=-\frac{\sqrt{2}}{3},~
\Braket{\frac{1}{2},1,\frac{3}{2}|(\boldsymbol L\cdot\boldsymbol s_2)|\frac{3}{2},1,\frac{3}{2}}=-\frac{\sqrt{5}}{3},
\end{equation}
where the 6j-symbols are taken from \cite{sobel}.

Using \eqref{eq:offdavg}, we obtain the leading order contributions to off-diagonal matrix elements of the Breit Hamiltonian \eqref{eq:eq1a}:
\begin{equation}
\label{eq:hfsoffdiag1}
E^{hfs,off-diag}_{F=1/2}=\frac{\alpha^4\mu^3(1+\kappa_{d})}{48m_1 m_2}\biggl( -\frac{\sqrt{2}}{6} \biggl)\biggl[ 1+\frac{2m_1 \kappa_{d}}{m_2(1+\kappa_{d})}-a_{\mu}\biggl]=-126.0372~\mu eV,
\end{equation}
\begin{equation}
\label{eq:hfsoffdiag2}
E^{hfs,off-diag}_{F=3/2}=\frac{\alpha^4\mu^3(1+\kappa_{d})}{48m_1 m_2}\biggl( -\frac{\sqrt{5}}{6} \biggl)\biggl[ 1+\frac{2m_1 \kappa_{d}}{m_2(1+\kappa_{d})}-a_{\mu}\biggl]=-199.2824~\mu eV.
\end{equation}
There exist higher order corrections to \eqref{eq:hfsoffdiag1} and \eqref{eq:hfsoffdiag2} which are related to additional interactions
and examined above.

\section{Relativistic corrections to hyperfine structure}
\label{app:fis1}

Relativistic corrections of order $\alpha^6$ can be calculated by means of the Dirac equation \cite{breit,Rose}.
We present here only a sketch of the output of the final formula for the numerical estimate.
In the Dirac theory the hyperfine part of relativistic Hamiltonian has the form:
\begin{equation}
\label{eq:relpot}
\Delta H^{hfs}=eg_N\mu_N\boldsymbol{s}_2\frac{[\boldsymbol{r}\times \boldsymbol{\alpha}]}{r^3},
\end{equation}
where $\mu_N$ is the nuclear magneton, $g_N$ is the deuteron gyromagnetic factor. To find the expectation value
of \eqref{eq:relpot} over atomic wave functions we should use the Wigner-Eckart theorem expressing initial
matrix element through the reduced matrix elements:
\begin{equation}
\label{eq:relgen}
\Delta E^{hfs}_{rel}=e g_N\mu_N (-1)^{s_2+j'-F}W(j s_2 j' s_2;F 1)\Braket{s_2||\boldsymbol{s_2}||s_2}\Braket{j'||\frac{[\boldsymbol{r}\times \boldsymbol{\alpha}]}{r^3}||j},
\end{equation}
Calculating the first reduced matrix element we can simplify
\eqref{eq:relgen} as follows:
\begin{equation}
\label{eq:relrel}
\Delta E^{hfs}_{rel}=e g_N\mu_N (-1)^{s_2+j'-F}\sqrt{(2s_2+1)(s_2+1)s_2} \sqrt{(2j'+1)(j'+1)j'}
W(j s_2 j' s_2;F 1)\times
\end{equation}
\begin{displaymath}
\times\Braket{j'\mu|\left(\frac{[\boldsymbol{r}\times\boldsymbol{\alpha}]}{r^3}\right)_z|j\mu}\mu^{-1}.
\end{displaymath}
In the case of diagonal matrix element we have:
\begin{equation}
\label{eq:akk}
\Braket{j\mu|\left(\frac{[\boldsymbol{r}\times \boldsymbol{\alpha}]}{r^3}\right)_z|j\mu}=-iA_{k k}R_{k k},~~~
R_{k k} = 2 \int^{\infty}_0 g_k(r) f_k(r) dr, ~~~-iA_{k k} = \frac{4k}{4k^2-1}.
\end{equation}
The radial matrix elements are calculated analytically with the use of exact Dirac radial wave functions.
After their expansion over $\alpha$ we find \cite{Rose}:
\begin{equation}
\label{eq:radp}
R(2P_{1/2})= \frac{(Z\alpha)^3}{12}\biggl(1+\frac{47}{24}(Z\alpha)^2\biggl)m_1^2,~~~
R(2P_{3/2})= -\frac{(Z\alpha)^3}{24}\biggl(1+\frac{7}{24}(Z\alpha)^2\biggl)m_1^2.
\end{equation}
As a result, general expressions for relativistic corrections to diagonal matrix elements take the form:
\begin{equation}
\label{eq:reldiag1}
E^{hfs}_{rel}(2P_{1/2})=\frac{\alpha^6(1+\kappa_d)\mu^3}{48 m_1 m_2}\frac{m_1^3}{\mu^3}\frac{47}{9}\times\frac{1}{2}[F(F+1)-J(J+1)-I(I+1)],
\end{equation}
\begin{equation}
\label{eq:reldiag2}
E^{hfs}_{rel}(2P_{3/2})=\frac{\alpha^6(1+\kappa_d)\mu^3}{48 m_1 m_2}\frac{m_1^3}{\mu^3}\frac{7}{45}\times\frac{1}{2}[F(F+1)-J(J+1)-I(I+1)].
\end{equation}
Numerical results for separate P-states are presented in Table~\ref{tb1}. Relativistic corrections to off-diagonal matrix elements
are evaluated in a similar way. The radial and angular integrals in this case take the form:
\begin{equation}
\label{eq:akk1}
R_{k k} = \int^{\infty}_0 \left[g_{\frac{1}{2}}(r) f_{\frac{3}{2}}(r)+g_{\frac{3}{2}}(r) f_{\frac{1}{2}}(r)\right] dr,~~~
-iA_l = \frac{[(l+1/2)^2-\mu^2]^{1/2}}{2l+1}=\frac{\sqrt{2}}{3},
\end{equation}
where the indexes near radial wave functions designate the values of muon total angular momentum $j$.
Radial integrations lead to analytical formulas and corresponding numerical results
\begin{equation}
\label{eq:reloffdiag1}
E^{hfs,off-diag}_{rel,F=1/2}=-\frac{\alpha^6(1+\kappa_d)\mu^3}{48 m_1 m_2}\frac{m_1^3}{\mu^3}\frac{3\sqrt{2}}{32}=-0.0043~\mu eV,
\end{equation}
\begin{equation}
\label{eq:reloffdiag2}
E^{hfs,off-diag}_{rel,F=3/2}=-\frac{\alpha^6(1+\kappa_d)\mu^3}{48 m_1 m_2}\frac{m_1^3}{\mu^3}\frac{3\sqrt{5}}{32}=-0.0067~\mu eV.
\end{equation}
Although their size is extremely small compared with other corrections we have included them in Table~\ref{tb2} by inserting
numerical values with an accuracy 0.0001 $\mu eV$ for the definiteness. It shows the relative numerical value of obtained corrections.

\section{Two-loop vacuum polarization corrections to hyperfine structure}
\label{app:fis1}

Two-loop vacuum polarization corrections presented in Fig.~\ref{fig:hfsVP1} b,c,d have the order $\alpha^6$.
We divide them into two parts: loop after loop contribution (vp-vp) and two-loop contribution to polarization operator
(2-loop vp). For their calculation we
use corresponding potentials in coordinate representation constructed in the same way as in \cite{apm2008}:
\begin{gather}
\notag\Delta V^{hfs}_{1\gamma,vp-vp}(r)=\frac{Z\alpha(1+\kappa_d)}{2m_1 m_2 r^3}\biggl( \frac{\alpha}{3\pi}\biggl)^2\int_1^\infty\rho(\xi)d\xi\int_1^\infty\rho(\eta)d\eta \frac{1}{\xi^2-\eta^2}\times\\
\label{eq:vpvppot}
\times\biggl[\biggl( 1+\frac{m_1\kappa_d}{m_2 (1+\kappa_d)} \biggl)(\boldsymbol L\cdot\boldsymbol s_2)
[\xi^2(1+2m_e \xi r)e^{-2m_e \xi r}-\eta^2(1+2m_e \eta r)e^{-2m_e \eta r}]-\\
\notag-(1+a_\mu)\biggl(4m_e^2 r^2[\xi^4 e^{-2 m_e \xi r}-\eta^4 e^{-2 m_e \eta r}]\times[(\boldsymbol s_1\cdot \boldsymbol s_2)-(\boldsymbol s_1\cdot\boldsymbol n)(\boldsymbol s_2\cdot \boldsymbol n)]+\\
\notag+[\xi^2(1+2m_e \xi r)e^{-2m_e \xi r}-\eta^2(1+2m_e \eta r)e^{-2m_e \eta r}]\times[(\boldsymbol s_1\cdot \boldsymbol s_2)-3(\boldsymbol s_1\cdot\boldsymbol n)(\boldsymbol s_2\cdot \boldsymbol n)]\biggl)\biggl],
\end{gather}
\begin{gather}
\notag\Delta V^{hfs}_{2-loop~vp}(r)=\frac{Z\alpha(1+\kappa_d)}{2m_1 m_2 r^3}\frac{2}{3}\biggl( \frac{\alpha}{\pi}\biggl)^2\int_0^1 \frac{f(v)dv}{1-v^2} e^{-\frac{2m_e r}{\sqrt{1-v^2}}}\times\\
\label{eq:2looppot}
\times\biggl[\biggl( 1+\frac{m_1\kappa_d}{m_2 (1+\kappa_d)} \biggl)\biggl[1+\frac{2m_e r}{\sqrt{1-v^2}} \biggl](\boldsymbol L\cdot\boldsymbol s_2)-\\
\notag-(1+a_\mu)\biggl(\frac{4m_e^2 r^2}{1-v^2}[(\boldsymbol s_1\cdot \boldsymbol s_2)-(\boldsymbol s_1\cdot\boldsymbol n)(\boldsymbol s_2\cdot \boldsymbol n)]+\\
\notag+\biggl(1+\frac{2m_e r}{\sqrt{1-v^2}} \biggl)[(\boldsymbol s_1\cdot \boldsymbol s_2)-3(\boldsymbol s_1\cdot\boldsymbol n)(\boldsymbol s_2\cdot \boldsymbol n)]\biggl)\biggl].
\end{gather}
Averaging \eqref{eq:vpvppot} and \eqref{eq:2looppot} over the Coulomb wave functions we obtain their numerical values in the hyperfine structure which are
presented in Table~\ref{tb1}-Table~\ref{tb2}. The muon vacuum polarization correction is evaluated by means of a replacement
$m_e\to m_1$ in \eqref{eq:1gammavp}. Its numerical value also is included in  Table~\ref{tb1}-Table~\ref{tb2}.

For the calculation of contributions in the second order PT we should use in basic expression
\begin{equation}
\label{eq:soptmainf}
\Delta E^{hfs}_{SOPT,vp}=2<\psi|\Delta V^{(1),C}_{vp}\cdot \tilde G\cdot\Delta
V^{(2),hfs}_{B,vp}|\psi>,
\end{equation}
the potential $\Delta V^{(2),hfs}_{B,vp}$ corresponding to pure hyperfine interaction or to hyperfine interaction
corrected by the vacuum polarization effect. Aa a second perturbation we use the Coulomb potential of one-loop or
two-loop order. All resulting matrix elements are calculated analytically in a standard way in the integration over the
coordinates of the particles and numerically by spectral parameters. Other details of their calculation can be founded in
our previous papers \cite{fmms,fmms1,apm2008}.

The two-loop vacuum polarization contribution to hyperfine structure of order $\alpha^6$ is determined also by the third order PT.
In this case we should use the following expression:
\begin{gather}
\notag\Delta E^{hfs}_{TOPT,vp}=\Braket{\psi_n|\Delta V^C_{VP}\cdot\tilde G \cdot \Delta V^{hfs}\cdot\tilde G \cdot \Delta V^C_{VP}|\psi_n}+\\
\notag+2\Braket{\psi_n|\Delta V^C_{VP}\cdot\tilde G \cdot V^C_{VP} \cdot\tilde G \cdot \Delta V^{hfs}|\psi_n}-\\
\notag-\Braket{\psi_n|\Delta V^{hfs}|\psi_n}\Braket{\psi_n|\Delta V^C_{VP}\cdot\tilde G \cdot\tilde G \cdot \Delta V^C_{VP}|\psi_n}-\\
\label{eq:TOPT}
-2\Braket{\psi_n|\Delta V^C_{VP}|\psi_n}\Braket{\psi_n|\Delta V^C_{VP}\cdot\tilde G \cdot\tilde G \cdot \Delta V^{hfs}|\psi_n}.
\end{gather}
Using further exact perturbation potential \eqref{eq:eq1a}, modification of the Coulomb potential $\Delta V^C$  and the Coulomb
Green's function $\tilde G$ \eqref{eq:green}, we obtain numerical values of corresponding corrections which are written in
Table~\ref{tb1} as a separate line. Numerically the vacuum polarization contributions of order $\alpha^6$ are extremely small
and will not have a significant impact on the comparison with future experimental data.

\end{document}